\documentstyle[12pt]{article}
\title{A formula with hypervolumes of six 4-simplices
and two discrete curvatures}
\author{Igor G. Korepanov\thanks{South Ural State University,
76 Lenin ave., Chelyabinsk 454080, Russia.
E-mail: kig@susu.ac.ru}}
\date{}

\def\be{\begin{equation}}
\def\ee{\end{equation}}
\def\ol{\overline}

\begin{document}
\maketitle

\begin{abstract}
One of the generalizations of the pentagon equation to higher dimensions
is the so-called ``six-term equation''. Geometrically, it corresponds
to one of the ``Alexander moves'', that is elementary rebuildings
of simplicial complexes, namely,
replacing a ``cluster'' of three 4-simplices by another ``cluster'', also
of three 4-simplices and with the same boundary. We present a formula
containing the euclidean volumes of the simplices in the first cluster in
its
l.h.s., and those in the second cluster --- in its r.h.s. The formula
also involves ``discrete curvatures'' appearing when we slightly deform
the euclidean space.
\end{abstract}

\bigskip
{\leftskip.3\textwidth \it
As an artist and scholar I prefer the specific detail to the generalization,
images to ideas, obscure facts  to  clear  symbols, and the discovered
wild fruit to the synthetic jam. \\[\smallskipamount] \null\hfill
Vladimir Nabokov

}

\section{Introduction}

This short note continues the paper~\cite{3s-t} and the short
note~\cite{5t}. We present a formula that naturally corresponds to
one of the ``Alexander moves''~\cite{alexander},
i.e., ``elementary rebuildings'' of simplicial complexes. Our formula
belongs to a four-dimensional space and deals with three ``initial''
4-simplices in its l.h.s. and three ``final'' ones in is r.h.s.

Recall that a similar formula for a three-dimensional space was obtained
in~\cite{5t}, and this was done on the basis of ``duality formulas''
(which are valid, themselves, for any-dimensional
space) from~\cite{3s-t}\footnote{The duality in~\cite{3s-t} can be said
to be dealing with ``branched polymers'' known in quantum gravity.}.

\section{Derivation of the formula}

Consider six points $A$, $B$, $C$, $D$, $E$ and $F$ in the four-dimensional
euclidean space\footnote{Yet, as will be seen below, we will allow them
sometimes to ``go out in the fifth dimension''.}. There exist fifteen
distances between them, which we denote, like in papers~\cite{3s-t,5t},
as $l_{AB}$, $l_{AC}$ and so on. There exist six 4-simplices with vertices
in our points, and we denote those simplices as $\ol A$, \ldots, $\ol F$,
where, say, $\ol A$ is the simplex $BCDEF$ ({\em not containing\/} the
vertex~$A$). The {\em four-dimensional hypervolume\/} we will denote as
$V$, e.g., $V_{\ol A}$ is the hypervolume of the simplex~$\ol A$.
We will need also the areas of {\em two-dimensional faces\/} ($S_{ABC}$
being the area of face $ABC$ and so on) and the ``defect angles'', or
``discrete curvatures'' concentrated in those faces (the defect angle
$\omega_{ABC}$ corresponds to the face $ABC$, etc.).

A defect angle means the following. With {\em arbitrary\/} distances
$l_{AB}$, \ldots, $l_{EF}$, the points $A$, \ldots, $F$ may not necessarily
be placed in the (``flat'') four-dimensional euclidean space.
Any {\em five\/} of those points, however, {\em can\/} be placed there,
thus forming a 4-simplex with vertices in those points, an then
one can calculate  the ``dihedral angles'' between its three-dimensional
hyperfaces. There are three such  ``dihedral angles'' at the
two-dimensional face $ABC$ --- they correspond to tetrahedra $\ol D$,
$\ol E$ and $\ol F$. In the flat case, the sum of those angles is $2\pi$,
and in the general case, it equals, by definition, $2\pi -\omega_{ABC}$.

All our considerations will take place in a small neighborhood of the
flat case $\omega_{ABC}=0$. Arguments perfectly analogous to those
in~\cite{5t}, but using \cite[formulas (15, 16)]{3s-t} instead of
\cite[formulas (11, 12)]{3s-t}, yield
\be
{1\over 12}\left| S_{ABC}\, l_{AB}\,dl_{AB}\over
V_{\ol D}\, V_{\ol E}\, V_{\ol F} \right| = \left| d\omega_{ABC}\over
V_{\ol A}\, V_{\ol B} \right|,
\label{64 eq 1}
\ee
if only $l_{AB}$ of all distances can vary. Similarly, one can write
\be
{1\over 12}\left| S_{DEF}\, l_{DE}\,dl_{DE}\over
V_{\ol A}\, V_{\ol B}\, V_{\ol C} \right| = \left| d\omega_{DEF}\over
V_{\ol D}\, V_{\ol E} \right|,
\label{64 eq 2}
\ee
if only $l_{AB}$ can vary.

If both $l_{AB}$ and $l_{DE}$ can change, but the zero curvature
is fixed:
\be
\omega_{ABC} \equiv 0,
\label{64 eq 3}
\ee
which is obviously equivalent to
\be
\omega_{DEF} \equiv 0,
\label{64 eq 4}
\ee
then $dl_{AB}$ and $dl_{DE}$ are related by
\be
\left| l_{AB} \,dl_{AB}\over V_{\ol D}\, V_{\ol E} \right| =
\left| l_{DE} \,dl_{DE}\over V_{\ol A}\, V_{\ol B} \right|
\label{64 eq 5}
\ee
(cf.~\cite[(16)]{3s-t}), and similarly one can write out the relation
between the differentials of any pair of distances.

Consider $\omega_{ABC}$ as a function of fifteen distances.
In a neighborhood of the flat configuration, we have
\be
d\omega_{ABC} = c_{AB} \,dl_{AB} + \cdots + c_{EF} \,dl_{EF},
\label{64 eq 6}
\ee
where all the ratios of coefficients $c_{\ldots}$ are fixed (at least,
up to a sign) by the formula~(\ref{64 eq 5}) and the like formulae
for other pairs of distances, e.g.,
\be
\left| c_{AB}\over c_{DE} \right| =
\left| l_{AB}\, V_{\ol A}\, V_{\ol B} \over l_{DE}\, V_{\ol D}\, V_{\ol E}
\right|,
\label{64 eq 7}
\ee
etc. If now we write, analogously,
\be
d\omega_{DEF} = c'_{AB} \,dl_{AB} + \cdots + c'_{EF} \,dl_{EF},
\label{64 eq 8}
\ee
then the coefficients $c'_{\ldots}$ will, obviously, have the {\em same\/}
ratios. Thus, the differentials of curvatures $\omega_{ABC}$ and
$\omega_{DEF}$ as functions of {\em all\/} distances are proportional,
namely, from (\ref{64 eq 1}, \ref{64 eq 2} and \ref{64 eq 5}) we find:
\be
\left| d\omega_{ABC}\over S_{ABC}\, V_{\ol A}\, V_{\ol B}\, V_{\ol C}
\right| =
\left| d\omega_{DEF}\over S_{DEF}\, V_{\ol D}\, V_{\ol E}\, V_{\ol F}
\right|.
\label{64 eq 9}
\ee

Let us, finally, write this in the form of the desired ``six-term
equation'':
\be
{S_{ABC}\, \delta(\omega_{ABC}) \over V_{\ol D}\, V_{\ol E}\, V_{\ol F} }=
{S_{DEF}\, \delta(\omega_{DEF}) \over V_{\ol A}\, V_{\ol B}\, V_{\ol C} }.
\label{64 eq 10}
\ee
Here $\delta$ is the Dirac delta function, and instead of writing out
the absolute value signs, we assume that the signs of (oriented)
hypervolumes and areas are chosen ``properly''. It is implied that both
sides of (\ref{64 eq 10}) can be integrated in any of $dl_{AB}$, \ldots,
$dl_{EF}$.

\section{Remarks}

\noindent {\bf 1.}\quad
Our equation (\ref{64 eq 10}) corresponds to a ``move of type
$3\to 3$'', i.e., three simplices are transformed into
three new ones. Nontrivial seems the question of what to do with
the other Alexander moves, that is $2\leftrightarrow 4$ and
$1\leftrightarrow 5$. Similarly, in the paper~\cite{5t} the moves
$2\leftrightarrow 3$ are analyzed, while $1\leftrightarrow 4$
requires further investigation.
\medskip

\noindent {\bf 2.}\quad
Our formulas are likely to be useful for quantum gravity. Namely, they
may help to find the most symmetric integration measure for ``functional
integrals'' in the discrete Regge-type models of space-time.
\medskip

\noindent {\bf 3.}\quad
The triangulated manifold where our rebuildings take place is not bound
to be flat --- see the similar Remark~2 in the end of paper~\cite{5t}.

\end{document}